\RequirePackage{fix-cm}
\documentclass[twocolumn,epjc3]{svjour3}  
\smartqed  
\RequirePackage{graphicx}

\usepackage{amsmath}
\def \pt {{p_\bot}}

\def \e {{\rm e}}
\def \eq {{\rm e}_q}
\def \d {{\rm d}}
\def \Q {{\bf Q}}
\def \q {{\bf q}}
\def \Z {{\cal Z}}
\def \Zq {{\cal Z}_q}
\def \p {{\cal P}}

\journalname{Eur. Phys. J. C}
\begin{document}

\title{Constrains for non-standard statistical models of particle creations by identified hadron multiplicity results 
at LHC energies}

\author{T. Wibig\thanksref{e1}}
\institute{Department of Physics, 
University of {\L }\'{o}d\'{z};  90-236 {\L }\'{o}d\'{z}, Pomorsta 149/153, Poland; \and
National Centre for Nuclear Research, Cosmic Ray Laboratory, \L \'{o}d\'{z}, Poland.}
\thankstext{e1}{e-mail: t.wibig@gmail.com}

\date{Received: date / Accepted: date}

\maketitle
\begin{abstract}
We analysed the identified hadron  multiplicity predictions of the modified
thermodynamical model of the multiparticle production
processes with non-exten\-sive statistic. The replacement of the standard Boltzmann exponential factor by the 
eventually much slower falling Tsallis one is suggested by the analysis of the transverse momentum distributions 
measured at high energies. The increase of high transverse momenta should accord with the abundance of heavy
secondary particles, in particular multistrange barions. The introduction to the thermodynamical model of suppression factors similar to the ones in a quark jet fragmentation models is discussed.
\end{abstract}

\section{Introduction}
The identified hadron ratios have been measured with all LHC detectors and results were compared with high-energy
event generators available in the market\cite{atlas,lhcb,aliceA,aliceE,aliceB}. The comparison, in general, is not very satisfactory. 

In the present paper we would like to use data from the ALICE experiment performed with $pp$ interaction of $\sqrt{s}$ 7 TeV available energy \cite{aliceA,aliceE,aliceB,omegadopi,aliceF,aliceG} to 
test the particle creation description based on thermodynamical approach. 

The standard statistical picture is known to work well in the soft, low $\pt$, sector of the particle creation process, where the
 exponential fall of the transverse momentum distribution is observed. The hard inelastic scattering leads to the quark jet fragmentation
with the power-law transverse momentum (transverse mass) distributions.
Detail studies of the measured charged particle transverse momentum (transverse mass) distributions suggested already some time ago
that the very good agreement of the invariant differential cross section in the whole transverse momentum range can be obtained with "an empirical formula inspired by QCD" from \cite{hage1983}  
\begin{equation}
\label{qcdinpired}
E~{ {\rm d ^3} \sigma \over  {\rm d} p ^3}  ~=~ A\: \left( \frac {p_0} {\pt+p_0} \right) ^{n}
\end{equation}
\noindent
 (see, e.g., \cite{wongwilk} for further discussion and references ).
It has been shown \cite{twkurp} that not only the fit of the simple form of Eq.(\ref{qcdinpired}) works well but the whole
theoretical model of particle creation which stands behind it could be successfully  applied to the highest available energy data on  charged particle transverse momentum \cite{jpg-q}.

The model parameters found in \cite{jpg-q} define the occupation 
of phase space for given charged particle transverse momentum. If the picture is self-consistent, 
the same set of parameters should give correct  yields of different kinds of created particles.     
It is known  that the multiplicities of new created heavy particles are described to some extent by the Boltzmann statistical model (e.g., \cite{redlich,bacatt97}). The Tsallis modification undoubtedly increases the high $\pt$s, and, obviously, the high transverse mass particle abundances. This should lead to the overabundance of heavy particles. We would like to look for the possibility to suppress this effect in a consistent way and to see if satisfactory results could be obtained.

\section{Thermodynamical model}
The thermodynamical picture of particle creation process in hadronic 
collisions was the first and quite successful attempt to describe it. 
The elaborated and complete theory was presented
in series of papers by Hagedorn (see \cite{hage} and references
therein). The idea of the fireball together with the proposition that 
``all fireballs are equal''  gives considerable predictions
concerning produced particle spectra. 

One of the predictions was that the temperature of the ``hadronic soup'' 
(precisely defined) could not exceed a universal
constant $T_0$ of order of 160 MeV. This value comes not as a result of the procedure of parameter adjusting using
multiparticle production (e.g., transverse momenta) data, but from examination of
elementary particle mass spectrum. 

The Hagedorn theory were abundant
for some time, when more sophisticated, jet or QCD based ideas appeared \cite{feynman}.
One of the reasons was the failure of the high-transverse momenta description.
The temperature of the fireball is defined as the parameter in the
classical Boltzmann exponential term of the probability weights for
phase space average occupation numbers. 
This gives the (asymptotic) form
of the distribution of transverse momentum of particles created from decaying
fireballs. It was found that at high and very high interaction energies 
the predicted exponential fall do not agree with the observed
high $p_\bot$ behaviour. 
Successes of QCD based description of the hard processes gave deep insight
into the nature of physics involved, and belief that this is just the
right theory of strong interactions, makes the thermodynamical approach
 very approximate, simple and naive tool of limited
applicability and thus of limited significance. But on the other hand, the
simplicity of the theory and notorious constant lack of the effective QCD
theory of soft hadronization processes give a hope that the fireball idea can
be enriched, modified and can become important again.

The Hagedorn idea was used again to describe the
identified particle multiplicities in hadronization both, in $e^+e^-$
annihilation and hadronic collisions. The {\it grand canonical} formalism of
Hagedorn was replaced in the serial of papers by Becattini and co-workers \cite{becattiniheinz}
by the {\it canonical} one, much relevant for studies of
small systems like primary created fireballs for which the requirement of
exact conservation of some quantum numbers seems important. 

In general,
thermodynamics of the system is determined by the partition function which can
be written as
\begin{equation}
\Z \left( Q^0 \right) ~=~\sum \limits_{ 
{Q}} \delta(Q-Q^0)
\: \prod \limits_{i,j} \p_{jk}^{\nu _{jk}} \ ,  
\label{z0}
\end{equation}

\noindent 
where $\p$ is the classical Boltzmann factor and $j$ and $k$ enumerate particle types and momentum
cells, $Q^0$ is the initial fireball quantum number vector and $Q$ is the
respective vector of the particular state, and $\nu_{jk}$ is the occupation
number. 
Introducing Fourier transform of $\delta$ (and reducing vector $Q$ to
3-dimensional: charge, baryon number and strangeness) Eq.(\ref{z0}) becomes 
\begin{eqnarray}
\Z\left(Q^0\right)~=~{1 \over (2 \pi)^3}\: 
\int \limits_{0}^{2\pi}
\int \limits_{0}^{2\pi}
\int \limits_{0}^{2\pi}
 {\rm d}^3  \phi
 \ {\rm e}^{iQ^0\phi} \times \ \ \ \ \ \ \ \ \ \ \ \ \ \ \ \nonumber \\ 
\ \ \ \ \ \ \ 
\times \exp \left\{
\sum _{j=1}^{n_B} \:w_j
\: \sum \limits_{k} \log \left( 1- \p_{jk} {\rm e}^{-iq_j\phi}\right)^{-1} \: 
\ + \ \right.\nonumber \\ + \left. \
\sum \limits_{j=1}^{n_F} \:w_j
\: \sum \limits_{k} \log \left( 1+ \p_{jk} {\rm
e}^{-iq_j\phi} \right) \:\right\} \label{z1}
 ,
\label{zq0}
\end{eqnarray}
where $q_{j}$ is the quantum number vector of  the particle $j$ and
$w_j$ is the weight factor associated  with the particle of the type $j$. The first guess is that it should be
equal to $(2 J_j + 1)$ and counts spin states. However, this does not seem to be so simple (see, e.g.,\cite{jetset}) and other solutions introducing factors responsible for some wave-function normalization, which
should disfavour heavier states
were found to be preferable by measurements. We will discuss this point later on. 

With the Eq.(\ref{zq0}) we are ready for detailed numerical calculations.

\subsection{Average multiplicities}

With the known partition function  $\Z$ the average
characteristics of the system can be obtained in an usual way. For the average multiplicity we have
\begin{eqnarray}
\langle n_j \rangle~=~
w_j\ { V \over \left(2\pi \right)^3 }\:
\:{1 \over \left(2\pi \right)^3}
\int \limits_{0}^{2\pi}
\int \limits_{0}^{2\pi}
\int \limits_{0}^{2\pi}
\d^3 \phi \: \times  \ \ \ \ \ \ \ \ \nonumber \\
\times \int \d ^3 p
\:\left[ \e^{E/T}\:\e^{i \:\q_j \phi} \pm 1 \right]^{-1}~,
\label{mult1}
\end{eqnarray}
where the upper sign is for fermions and the lower is for bosons.
Because the $\e^{-E/T}$ factor is expected to be small (for 
all particles except pions)
then
\begin{eqnarray}
\langle n_j \rangle~
\approx~ 
{\Z(\Q^0 -\q_j) \over \Z(\Q^0)}
\:w_j
{ V \over \left(2\pi \right)^3 }\:
\int \d ^3 p
\:\e^{-E/T}~.
\label{mult2}
\end{eqnarray}

The conventional Boltzmann-Gibbs
description shown above could be, in principle, modified  to allow the 
description of the systems of not-completely-free particles: the correlation
``strength'', however defined, was introduced with the help of the new
{\it non-extensivity}  parameter
and the new statistics, which in such case has to be also {\it non-extensive}. 
In the limit of the absence of correlations the 
new description approaches the Boltzmann form.

There could be
infinitely many ``generalized'' statistics which fulfill such requirements.
We choose
the one which is simple and has well defined 
theoretical background. In the present paper we
test the possibility, proposed by Tsallis \cite{Tsallis:1988eu},
based on the modification of the 
classical entropy definition
\begin{equation}
S_{\rm BG}~=~-k \sum \limits_{i}^{W} \:\p_i\ln \p_i
\label{entrobg}
\end{equation}
by the new one
\begin{equation}
S_{q}~=~k ~{1 \over q~-1}\left({1-\sum \limits_{i}^{W} \:\p_i^q }\right)~
\label{entrots}
\end{equation}
with the new parameter $q$ called the non-extensivity parameter.
This modification has been adopted in
other physical applications (see, e.g., \cite{beck}).

Maximization of the entropy requirement with the 
total energy constraint
leads to the probability of realization of the state $i$ (with energy $E_i$) given by

\begin{equation}
\p_i~=~{1\over \Z_q}\:
\left[1-(1-q)/T_q(E_i-E_0) \right]^{1/(1-q)}~,
\label{peq0}
\end{equation}
where $\Z_q$ is the normalization constant related to $\Z(\Q^0)$
of Eq.(\ref{z0}) where the Boltzmann
terms $x$ is replaced by the probabilities of the form of Eq.(\ref{peq0}).

Eq. (\ref{peq0}) can be rewritten introducing new
symbol, $\eq$, defined as $
\eq^{x}  =
\left[ 1+(1-q)x\right]^{1/(1-q)}
$
\begin{equation}
\p_i~\sim~
\eq ^{ -E_i/T_q}
\label{peq}
\end{equation}
and the modified partition function can be written then in the form
\begin{equation}
\Zq(\Q)=\sum \limits_{\rm states}
{w_j \over \left(2\pi \right)^3}
\int \limits_{0}^{2\pi}
\int \limits_{0}^{2\pi}
\int \limits_{0}^{2\pi}
\d^3 \phi \:
\eq^{-E/T} \:
\e^{i (\Q_0-\Q)\:\phi} .
\label{zqdef}
\end{equation}
Eq.(\ref{mult2}) with this modification of the partition function gives abundances of initially
created particles in the hadronization process described by the modified, non-extensive, statistics.

\section{Results}
We have evaluated $\Zq$ functions (and $\Z=\Z_1$ for the standard Boltzmann thermodynamics) for a variety of values 
of the thermodynamical model parameters: 
$T$, $V$, and a number of $\Q$ values which cover the production of over 100 hadrons 
of masses below 2 GeV/c$^2$. All decays of short-lived particles were then taken into account. 

Measurements of identified particle ratios performed mostly by  the ALICE Collaboration 
give the opportunity to test the modified statistical model of particle flavour creation in the new, 
higher energy region. 
The lower energy jet hadronization results have been analyzed by the serial of papers by Becattini 
and others (see, e.g., \cite{redlich,bacatt97}). It has been shown that the micro-canonical Boltzmann 
description works well for $e^+e^-$ from $\sqrt{s}\approx $10 GeV \cite{becatpesa} to 91 GeV 
\cite{redlich} and $p p$ and $p \bar p$ interactions up to SPS evergies,  
$\sqrt{s}\approx $ 900 GeV \cite{bacatt97}.

\begin{figure}
 \includegraphics[width=7.3cm]{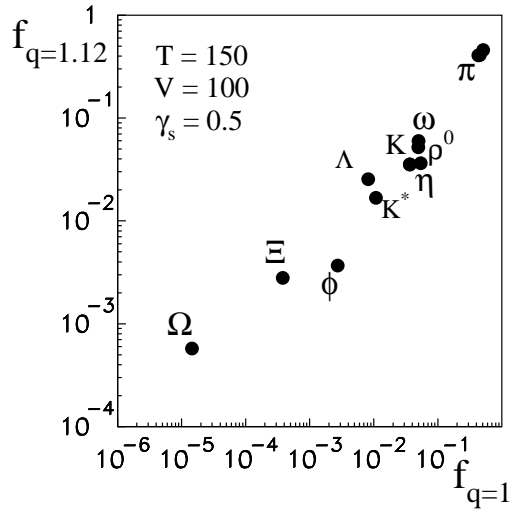} 
\caption{Relative particle multiplicities, $f = \left( n_i/n_{ch}\right)$, obtained for 
the Tsallis statistics (with $T=150$ MeV, $V=100$ fm$^3$ and $q=1.12$) compared with the same ratios 
for the Boltzmann statistics (with $T=150$ MeV, $V=100$ fm$^3$ and $q=1$).}
\label{f1}
\end{figure}

\begin{figure}
 \includegraphics[width=7.3cm]{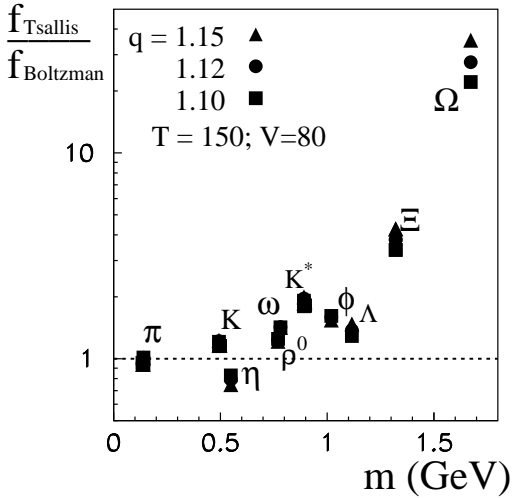} 
\caption{Enhancement of the relative particle multiplicity obtained for Tsallis statistics (with $T=150$ MeV, $V=80$ fm$^3$) with respect to the standard Boltzmann model for the non-extensivity parameter values equal to 1.10,1.12, and 1.15 (squares, circles and triangles, respectively).}
\label{f2}
\end{figure}

The comparison of the results obtained with Boltzmann (non-extensivity parameter $q=1$) 
and Tsallis with the value of $q = 1.12$ which obtained in Ref.~\cite{twkurp,jpg-q} is shown 
in Fig.~\ref{f1}. We can see the clear enhancement of the exponential 'tail' for modified statistics model. 
The non-extensivity parameter $q$ values adjusted to high energy data recently \cite{cleymans}
are of 1.1--1.15,  ($\sim1.17$ in Ref.~\cite{wongwilk}). We have shown the effect of the change of
the non-extensivity parameter in Fig.~\ref{f2}. The difference between the Boltzman and Tsallis 
statistics results are given for three values of $q$,
 1.10, 1.12, and 1.15. 
The biggest  difference is seen for $\Omega$ . For lighter particles, even for $\Xi$ the effect of small change of $q$ is not
significant. Concluding, we can say that the relative particle multiplicities are not the appropriate observable
to adjust the non-extensivity parameter.

We have tested also the particle multiplicity dependencies on thermodynamical hadronization parameters $T$
and $V$. 
The effect seems to be almost negligible with the limits possible changes, allowed by the data on transverse momentum distribution and total multiplicities. However, 
the effect of volume $V$ have to be taken with care, because the spacial and temporal history of 
the hadronization process is not exactly known. It is expected that the canonical picture should 
take into account the multichain idea (e.g., \cite{twgmc,becattiniheinz}) of decomposition of 
the 'hadronic soup' to chains of independently hadronized objects/fireballs. For the Boltzmann
statistics, by the definition of extensivity, the sum of many hadron sources is equivalent to one 
big source \cite{becatpesa}. This is, in general, not the case of Tsallis non-extensive statistics. 
But we can say that the strength of the non-extensivity is still not big and the effect of
subdivision of the hadronization volume does not change much the conclusion about the identified 
particles ratios. Another important point, to be mentioned here, is connected to the effect of the canonical treatment of small 
fireballs which relates to the suppression of strange quark (and diquark, or strange diquark) production what 
was mentioned already in \cite{hage}. Additionally the importance of reaction volume in hadronic 
collisions in the canonical picture, specially for multistrange particles, like $\Omega$, 
is discussed extensively in \cite{rafelski}. 
We have to say that the changes of the hadronization volume do not act strong on the total
multiplicities. The possible small changes which we can study do not effect the 
particle ratios in a significant way. Detailed studies, however, are needed to answer all questions here.

We can say, that the thermodynamical model predictions are, in a sense, very robust. They
cannot be adjusted to the measured ratios, at least with reasonable possible changes of the hadronization parameters $T$, $V$ and $q$. This situation is, on the other side, very fortunate.
The comparison of them with the experimental results could be the {\em experimentum crucis} 
of the model in general. 

We have shown results for particle ratios in the comparison with
the ALICE Collaboration data in Tab.~\ref{t1}. 
Thermodynamical parameter values ($T$ and $q$) were taken from the literature, and adjusted ($V$) to 
reproduce roughly the charged particle multiplicities. We have applied there the simply counting of spin states to calculate the
weight factor $w_i$, $(2J_i+1)$, and strangeness suppression factor value of $\gamma_s = 0.5$ acting of strange quark particle contents.

\begin{center}
\begin{table*}[ht]
\caption{Ratios of particle multiplicities calculated for Boltzmann and Tsallis statistics  in comparison with results from the ALICE Collaboration. \label{t1}}

\begin{center}\begin{tabular}{|cc|ccc|}
\hline
particle ratio& ALICE measurement& $\begin{smallmatrix}\\ Boltzmann \\ \\ V=50\ {\rm fm}^3 \\T=160\ {\rm MeV}\\q=1.00\\ \\ \end{smallmatrix}$ &
 $\begin{smallmatrix}\\ Tsallis \\ \\ V=80\ {\rm fm}^3\\T=170\ {\rm MeV} \\q=1.12\\ \\ \end{smallmatrix}$ &
 $\begin{smallmatrix}\\ Tsallis \\ \\ V=80\ {\rm fm}^3\\T=150\ {\rm MeV} \\q=1.15\\ \\ \end{smallmatrix}$ \\
\hline
$\rho / \omega$ & $1.15 \pm 0.2 \pm 0.12^a$\footnotetext{$^a$1 GeV/c $< p_\bot<5$ GeV/c \cite{aliceF}}         &0.985& 0.855&0.848\\
$\phi /(\rho + \omega) $&$0.084 \pm 0.013  \pm 0.012^a$                                  &0.042  &0.035&0.033 \\
$K^{*0} / K^- $ &$0.35 \pm 0.001 \pm 0.04^b$\footnotetext{$^b$full phase space \cite{aliceA}}  &0.337 &0.466&0.466  \\
$\phi / K^{*0} $&$0.33 \pm 0.004 \pm 0.05^b$                                                                 &0.268   &0.215&0.207 \\
$\phi / \pi^{-}$&$0.014 \pm 0.0002\pm 0.002^b$                                                              &0.0063 &0.0080&0.0077 \\
$\phi / K^{-}$&$0.11 \pm 0.001 \pm 0.02^b$                                                                     &0.090&  0.100 & 0.097      \\
$\omega/\pi^0$&$  0.6 \pm 0.1^c$ \footnotetext{$^c  p_\bot > 2.5$ GeV/c \cite{omegadopi}}                         & 1.36&  0.861 &0.704   \\
%$\Xi / \pi$ &$ 0.0035\pm 0.0005$                                       & 0.0119&0.0110  &0.00948\\
%$\bar p / \pi^{-}$ &$.055 \pm 0.003$&                                    0.138 &.214  &0.148\\
$\Omega /\Xi$&$0.067\pm 0.01^d$\footnotetext{$^d m_\bot - m_0>0.3$ GeV \cite{aliceB}}                     & 0.068 &0.237  &0.240\\
$\Omega /\phi$ &$0.04 \pm .008^e$\footnotetext{$^e p_\bot>1$ GeV \cite{aliceA}}                                 & 0.119 &0.362     &0.403\\
$\eta / \pi^0$ &$0.1067 \pm 0.0259 \pm 0.0212^f$\footnotetext{$^f p_\bot>0.55$ GeV/c \cite{aliceE}}     & 0.206  &0.092  &0.081\\
\hline
\multicolumn{4}{l}{\scriptsize $^a$1 GeV/c $< p_\bot<5$ GeV/c \cite{aliceF}} \\
\multicolumn{4}{l}{\scriptsize$^b$full phase space \cite{aliceA}}\\
\multicolumn{4}{l}{\scriptsize$^c  p_\bot > 2.5$ GeV/c \cite{omegadopi}}\\
\multicolumn{4}{l}{\scriptsize$^d m_\bot - m_0>0.3$ GeV \cite{aliceB}}\\
\multicolumn{4}{l}{\scriptsize$^e p_\bot>1$ GeV \cite{aliceA}}\\
\multicolumn{4}{l}{\scriptsize$^f p_\bot>0.55$ GeV/c \cite{aliceE}}\\ 
\end{tabular}

\end{center}\end{table*}
\end{center}

Problems when comparing the ALICE results with prediction of listed hadronisation model results can be seen.
%We can list here $\chi^2$ values for  the data and three models  in the Table \ref{t1}: 170, 2000, 2500, respectively.  
Some ratios, especially those involving strange and 
multi-strange hadrons looks unexplainable.
As it is discussed above (Fig.~\ref{f2}) formal fits or readjustments of the model parameters ($T$, $V$ and even $q$) could not help. It should be mentioned here again, that model parameters are related to other interaction properties measured extensively with LHC, e.g.,  the transverse momentum distributions and total multiplicities, and their values are rather fixed. Any significant change of $V$, $T$ (and $T$ with $q$) could disturb fits made for the charged particle inclusive spectra measured with very high accuracy and in a large 
range of the transverse momentum space.

There is, however, in Eqs.~(\ref{mult1},\ref{mult2})
the weight factor $w_j$  which give us some 
hope and freedom to get closer to the data. The simply obvious form of $(2J+1)$ is, in general, modified since 
there had been experimentally found suppression of $K$ mesons with respect to non-strange ones. 
The general statement is that the strange phase space is not fully available for particle production what 
can be realized by multiplying the partition function by the special factor for each strange valence quark 
in the particle in question. 

The strangeness suppression factor is also one of the basic parameters in the jet fragmentation 
model introduced by Feynman and developed finally by the Lund group \cite{jetset}.
In the Lund jet fragmentation process new hadrons appear as a breaking of the colour field string 
stretched between quarks moving apart by the production of a new pair of quarks (sometimes diquarks). 
If there is enough energy left, further breaks may occur and eventually only on-mass-shell hadrons remains. 
The creation of new quark-antiquark pair in the Lund model is a kind of the quantum tunneling process, 
so it is expected that the heavy quark creation is suppressed. 
It is usually assumed that  $u : d : s \sim 1 : 1 : 0.3$ \cite{wroblewski}
Additionally, $w_j$ weight factor is related with the spin states of newly created hadrons: for mesons: 
pseudoscalar and vector states. The  suppression here is not defined in the Lund fragmentation 
model. Counting the spin states gives $1 : 3$ ratio, but in the JETSET model this ratio is eventually close to $1:1$, according to
the 'tunnelling normalisation'.

The situation with baryon creation in the Lund model is much more complicated. The tunneling mechanism 
is also adopted here. We have here the probability of string breakup via diquark mode and further 
combination of quark and diquark. If we take into account the pop-corn mechanism 
of diquark breakups and lack of general rules, we can have the number of parameters to be adjusted to 
the data comparable with the number of measured ratios to be used for this adjustment. 
The number of parameters describing the production of baryons measured with good accuracy in the 
experiments at LHC is at the moment higher then the number of such baryons itself \cite{jetset}.
 
The Lund model and particular JETSET hadronization generator is used also by the PHOJET \cite{phojet} 
program package for the recent theoretical examination of the LHC data description and comparison. Some
parameters in PHOJET are different than the default Lund model values.

We first discuss the possibility of introducing the strangeness suppression factor $\gamma_s$ for 
\begin{equation}
w_j~ =~ (2 J\:+\:1) \times \left( \gamma_s\right)^{N_j}~~~~. 
\label{eqwjgamma}
\end{equation}
The $N_j$ is the 'degree of strangeness' which is, in fact, not defined, yet. 
It should be related to the contents of the particle $j$. Three possibilities are rather natural.

\begin{equation}
N_j~=~\left\{
{
\begin{array}{lr}
S&~~{\rm  strangeness~ of ~the~ particle~of~ the~ type~ }j\\
n_s& ~{\rm  number~ of~ strange~ (or~ anti strange)}~~~~~~~ \\
~~ &~~~~~~~~~~~~ {\rm valence~ quarks~ of~}j\\
n_{s \bar s}&{\rm number~ of~ }s \bar s~{\rm  pairs~~~~~~~~~~~~~~~~~~~~~~~~~~~~}\\
& ~~~~~~{\rm  involved~ to~ create~the ~ particle~ }j\\
\end{array}
}
\right.
\label{eq16}
\end{equation}
\noindent
The difference could be seen in the comparison of $K$ and $\phi$ weights. For the direct $K$ they are 
$\gamma_s$, $\gamma_s$, and $\gamma_s$, for three possibilities in Eq.(\ref{eq16}), respectively. While for the direct $\phi$ they are 1, $\gamma_s^2$, and $\gamma_s$, 
for the first, the second and the third possibility in Eq.(\ref{eq16}). The actual situation is more complicated, 
because of the effect of decays of heavy resonances. To see the eventual results complete calculations have to 
be performed.

Some examples are given in the Tab.~\ref{t2}. We show there only these ratios which are sensitive to the choice of $N_j$
 in Eq.~(eq16).

\begin{center}
\begin{table*}[ht]
\caption{Some ratios of particle multiplicities calculated with $T=150\ {\rm MeV}$, $V=80\ {\rm fm}^3$, $q=1.12$ and 
different strangeness suppression factor definition in Eq.(\ref{eq16}) (the strangeness suppression factor equal to 
$\gamma_s = 0.5$) (description like in Tab.\ref{t1}).\label{t2}}

\begin{center}\begin{tabular}{|cc|ccc|}
\hline
particle ratio& ALICE measured ratios&$n_{s}$& $n_{s \bar s}$& $S$\\
\hline
%$\rho / \omega$&$1.1 \pm 0.2 \pm 0.12^a$%\footnotetext{$^a$1 GeV/c $< p_\bot<5$ GeV/c}     
%&0.862   & 0.870 &0.888\\
%$\phi /(\rho + \omega) $&$0.416 \pm 0.032  \pm 0.04^a$              
$\phi /(\rho + \omega) $&$0.084 \pm 0.013  \pm 0.012^a$                         &0.0360  &0.0718 &0.142 \\
%$K^{*0} / K^- $ &$0.35 \pm 0.001 \pm 0.04^b$%\footnotetext{$^b$full phase space}           
%&0.462   &0.423  &0.382  \\
$\phi / K^{*0} $&$0.33 \pm 0.004 \pm 0.05^b$                                    &0.219   &0.440  &0.751 \\
$\phi / \pi^{-}$&$0.014 \pm 0.0002\pm 0.002^b$                                   &0.008  &0.016 &0.030\\
$\phi / K^{-}$&$0.11 \pm 0.001 \pm 0.02^b$                                      &0.101   &0.186  &0.287   \\
%$\omega/\pi^0$&$  0.6 \pm 0.1^c$% \footnotetext{$^c  p_\bot > 2.5$ GeV/c}                  
%& 0.894  & 0.893 & 0.870  \\
%$\Omega /\Xi$&$0.067\pm 0.01^d$%\footnotetext{$^d m_\bot - m_0>0.3$ GeV}                  
% & 0.233  &0.219  &0.233  \\
$\Omega /\phi$ &$0.04 \pm .008^e$%\footnotetext{$^e p_\bot>1$ GeV}                        
 & 0.329  &0.135  & 0.082   \\
%$(K^{+}+K^{-}) \over (\pi^{+}+\pi^{-})$ &$0.159 \pm .005^f$%\footnotetext{$^f p_\bot>0.5$ GeV \cite{aliceG}} 
%&                                      0.115  &0.126    &0.158\\
%$\eta / \pi^0$ &$0.1067 \pm 0.0259 \pm 0.0212^f$%\footnotetext{$^{a,b,c,d,e,f}$ - like in Tab.I}
%\footnotetext{$^f p_\bot>0.55$ GeV/c}     
%& 0.100  &0.102  &0.103  \\
\hline

\end{tabular}
\end{center}\end{table*}
\end{center}

It can be seen that calculated ratios shown in Tab.~\ref{t2} that the case ($N_j=n_{s}$) 
works well for strange mesons and the simple $(2J+1)$ factor results not far from measurements. 
The multiplicity of $\phi$ is crucial here, as it could be expected.

Further model 'fine tuning' can involve the  adjustment of the value of the strangeness suppression 
factor $\gamma_s$. We did not, however, go very far. We have checked 
three values which have been used in the literature: 0.5 originally proposed by Feynman in the jet 
fragmentation model \cite{feynman} and still of use \cite{becattiniheinz}, 2/3 used 
successfully by Becattini \cite{becattni066} for the $e^+e^-$ data and $p \bar p$ results from SPS, 
and $\gamma_s=1$ as the limit of no strangeness suppression.  
Some results  for $n_{s}$ choice in Eq.(\ref{eq16}) and spin counting states  $w_i=(2J_i+1)$  are shown in Tab.\ref{t3} ($T=150\ {\rm MeV}$, $V=80\ {\rm fm}^3$, $q=1.12$).
\begin{center}
\begin{table*}[ht]
\caption{Some ratios of particle multiplicities calculated for and 
different values of strangeness suppression factor compared with the measurement results.\label{t3}}

\begin{center}\begin{tabular}{|cc|ccc|}
\hline
particle ratio& ALICE results &$\gamma_s=1$& $\gamma_s=2/3$& $\gamma_s=1/2$\\
\hline
%$\rho / \omega$ &$1.15\pm0.2\pm 0.12^a$%\footnotetext{$^a$1 GeV/c $< p_\bot<5$ GeV/c}
 %                                                                           &0.882  & 0.867  &0.861\\
$\phi /(\rho + \omega) $&$0.084 \pm 0.013  \pm 0.012^a$                              &0.137  &0.062  &0.036 \\
%$K^{*0} / K^- $ &$0.35 \pm 0.001 \pm 0.04^b$%\footnotetext{$^b$full phase space} 
%                                                                            &0.429  &0.453   &0.462  \\
$\phi / K^{*0} $&$0.33 \pm 0.004 \pm 0.05^b$                                           &0.433 &0.289  &0.219 \\
$\phi / \pi^{-}$&$0.014 \pm 0.0002\pm 0.002^b$                                         &0.023&0.013&0.008\\
$\phi / K^{-}$&$0.11 \pm 0.001 \pm 0.02^b$                                             &0.186&0.131&  0.101   \\
$\omega/\pi^0$&$  0.6 \pm 0.1^c$ %\footnotetext{$^c  p_\bot > 2.5$ GeV/c}
               & 0.787& 0.857  & 0.894  \\
$\Omega /\Xi$&$0.067\pm 0.01^d$%\footnotetext{$^d m_\bot - m_0>0.3$ GeV}
                & 0.443 &0.303&0.233  \\
$\Omega /\phi$ &$0.04 \pm .008^e$%\footnotetext{$^e p_\bot>1$ GeV}
                      & 0.585 &0.416& 0.329     \\
%$(K^{+}+K^{-}) \over (\pi^{+}+\pi^{-})$ &$0.159 \pm .005^f$%\footnotetext{$^f p_\bot>0.5$ GeV \cite{aliceG}} 
%&                                      0.195  &0.142    &0.115\\
$\eta / \pi^0$ &$0.1067 \pm 0.0259 \pm 0.0212^f$%\footnotetext{$^{a,b,c,d,e,f}$ - like in Tab.I}
%\footnotetext{$^f p_\bot>0.55$ GeV/c}  
                                                                           & 0.084  &0.094&0.161  \\
\hline
\end{tabular}
\end{center}\end{table*}
\end{center}
The  agreement for $\gamma_s=2/3$ seems slightly better than for 1/2. The no-suppression ($\gamma_s=1$)
is in general worst.

The discrepancy exists still in ratios 
involving baryons $\Omega$ and $\Xi$.
As it has been said there is a great degree 
of freedom to modify created baryon multiplicities. The diquark suppression factor $\gamma_{qq}$ is the 
one possibility which we used and another factor was introduced specially for $\Omega$ 
baryons $\gamma_{ss}$ related to creation of double strange diquark. 

Our final results on ratios of particle multiplicities calculated for with 
strangeness suppression factor $\gamma_s=2/3$ and extra diquark suppressions $\gamma_{qq}=\gamma_{ss}=1/2$
 are shown in the Tab.~\ref{t4} ($T=150\ {\rm MeV}$, $V=80\ {\rm fm}^3$ and $q=1.12$).

\begin{center}
\begin{table}[ht]
\caption{Ratios of particle multiplicities calculated for with 
different strangeness suppression factor $\gamma_s=0.66$ and extra diquark suppressions $\gamma_{qq}=\gamma_{ss}=0.5$ compared with the measurement results (description like in Tab.\ref{t1}).\label{t4}}

\begin{center}\begin{tabular}{|cc|c|}
\hline
particle ratio& ALICE results&calculated\\
\hline
$\rho / \omega$ 
& 1.15 $\pm$ 0.2 $\pm 0.12^a$%\footnotetext{$^a$1 GeV/c $< p_\bot<5$ GeV/c}   
          &0.867 \\
$\phi /(\rho + \omega) $&$0.084 \pm 0.013  \pm 0.012^a$                      &0.062\\
$K^{*0} / K^- $ &$0.35 \pm 0.001 \pm 0.04^b$%\footnotetext{$^b$full phase space}
        &0.453\\
$\phi / K^{*0} $&$0.33 \pm 0.004 \pm 0.05^b$                                            &0.302\\
$\phi / \pi^{-}$&$0.014 \pm 0.0002\pm 0.002^b$                                        &0.0147\\
$\phi / K^{-}$&$0.11 \pm 0.001 \pm 0.02^b$                                             &0.136 \\
$\omega/\pi^0$&$  0.6 \pm 0.1^c$ %\footnotetext{$^c  p_\bot > 2.5$ GeV/c}
              & 0.872\\
$\Omega /\Xi$&$0.067\pm 0.01^d$%\footnotetext{$^d m_\bot - m_0>0.3$ GeV}     
          &0.103\\
$\Omega /\phi$ &$0.04 \pm .008^e$%\footnotetext{$^e p_\bot>1$ GeV} 
                     & 0.045 \\
%$(K^{+}+K^{-}) \over (\pi^{+}+\pi^{-})$ &$0.159 \pm .005^f$%\footnotetext{$^f p_\bot>0.5$ GeV \cite{aliceG}} 
%&                                      0.124     &0.153\\
$\eta / \pi^0$ &$0.1067 \pm 0.0259 \pm 0.0212^f$%\footnotetext{$^{a,b,c,d,e,f,g}$ - like in Tab.I}
 &0.105 \\
\hline
\end{tabular}
\end{center}
\end{table}
\end{center}

Because of relatively limited amount of data we do not wish at the moment to go further 
with 'tuning' the suppression parameters ($\gamma_s$, $\gamma_{qq}$ and $\gamma _{ss}$). We would like to show
the general possibility to improve the data description in the thermodynamical model by introducing diquark suppression factors. 
The additional suppression of heavy, strange barions required by the modified thermodynamical model can be naturally 
realized this way.

With all the modifications described above the $\chi^2$ for the values listed in Tab. \ref{t4} gets lower from enormous 
thousands for predictions shown in Tab. \ref{t1} results to about 30. 
This value has a chance probability of $p=0.0004$, equivalent to '3.5 $\sigma$' deviation. 
It is, in fact, the disagreement, but it gives also a hope to be reduced further with more sophisticated calculations
and model improvements.

\begin{figure*}
\centerline{
 \includegraphics[width=7.3cm]{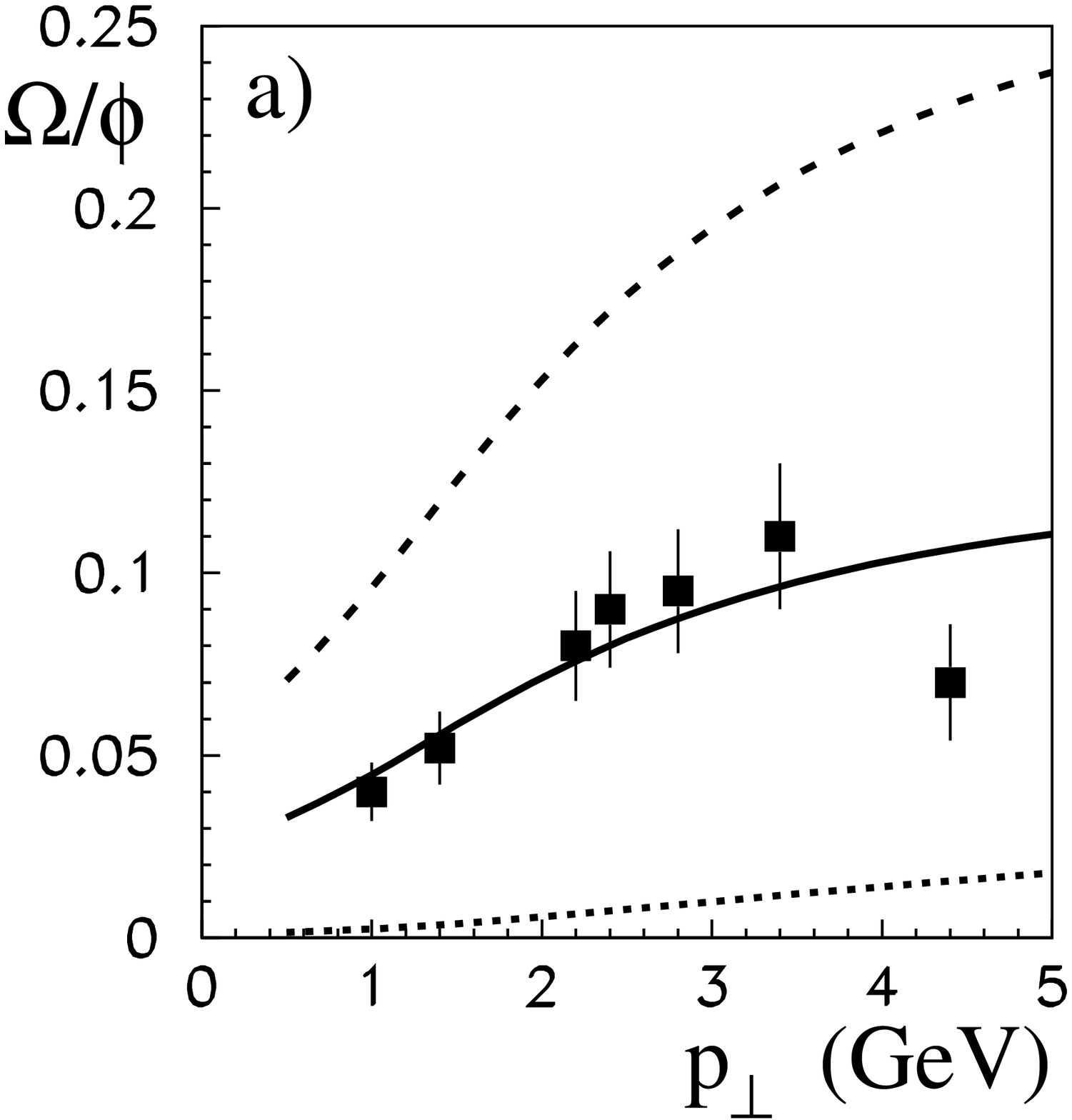} 
 \includegraphics[width=7.3cm]{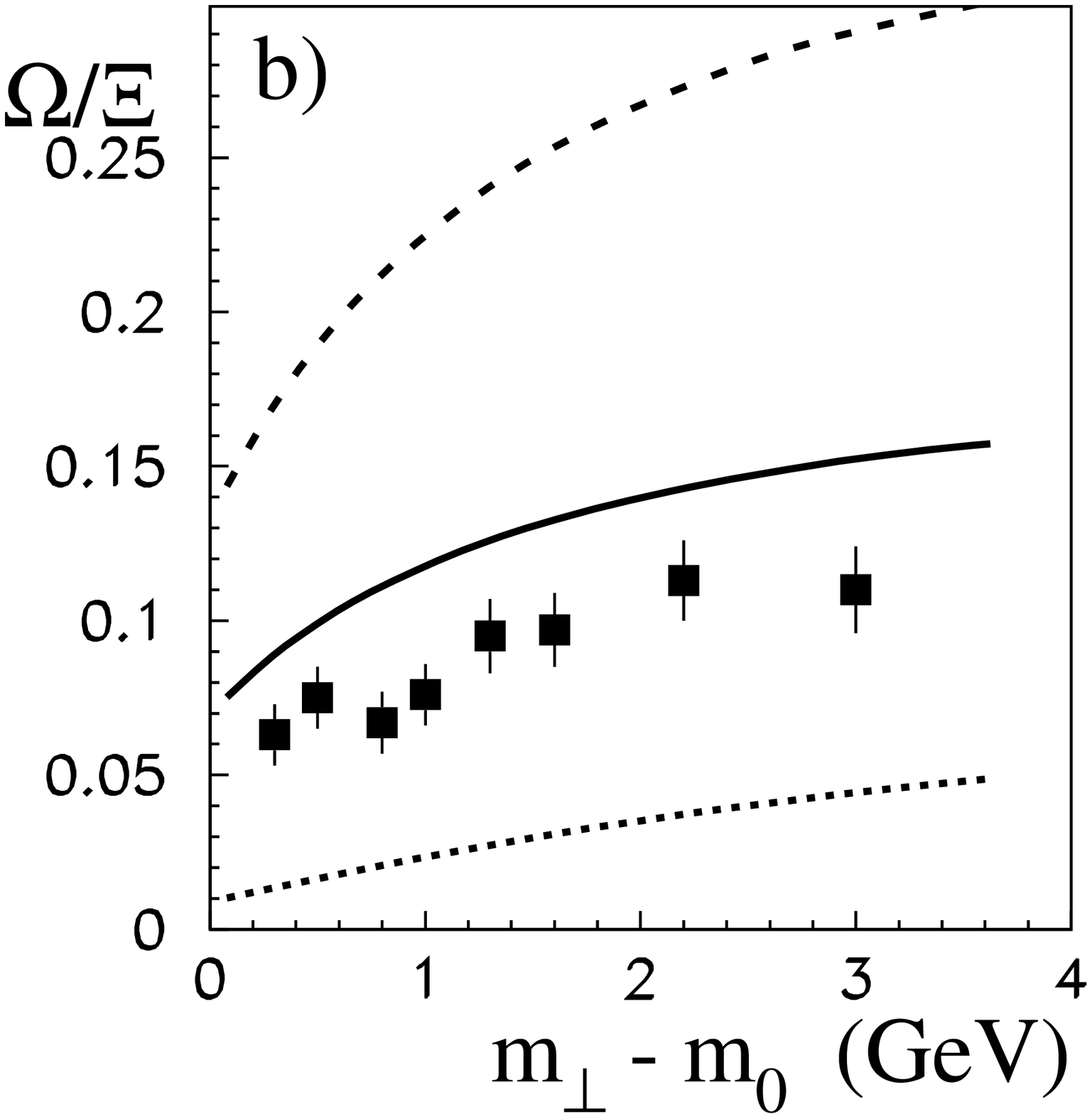} }
\centerline{
 \includegraphics[width=7.3cm]{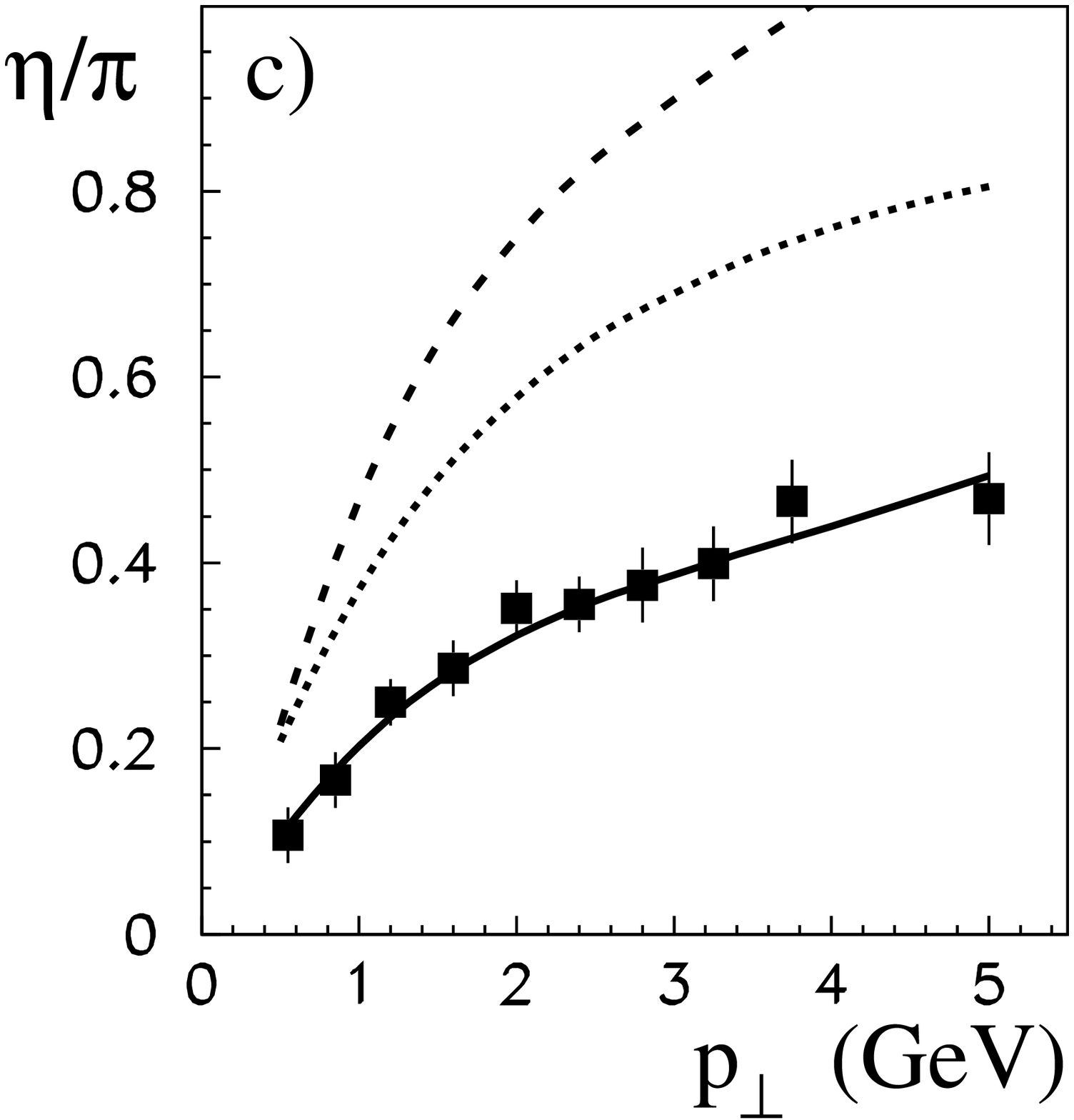} 
 \includegraphics[width=7.3cm]{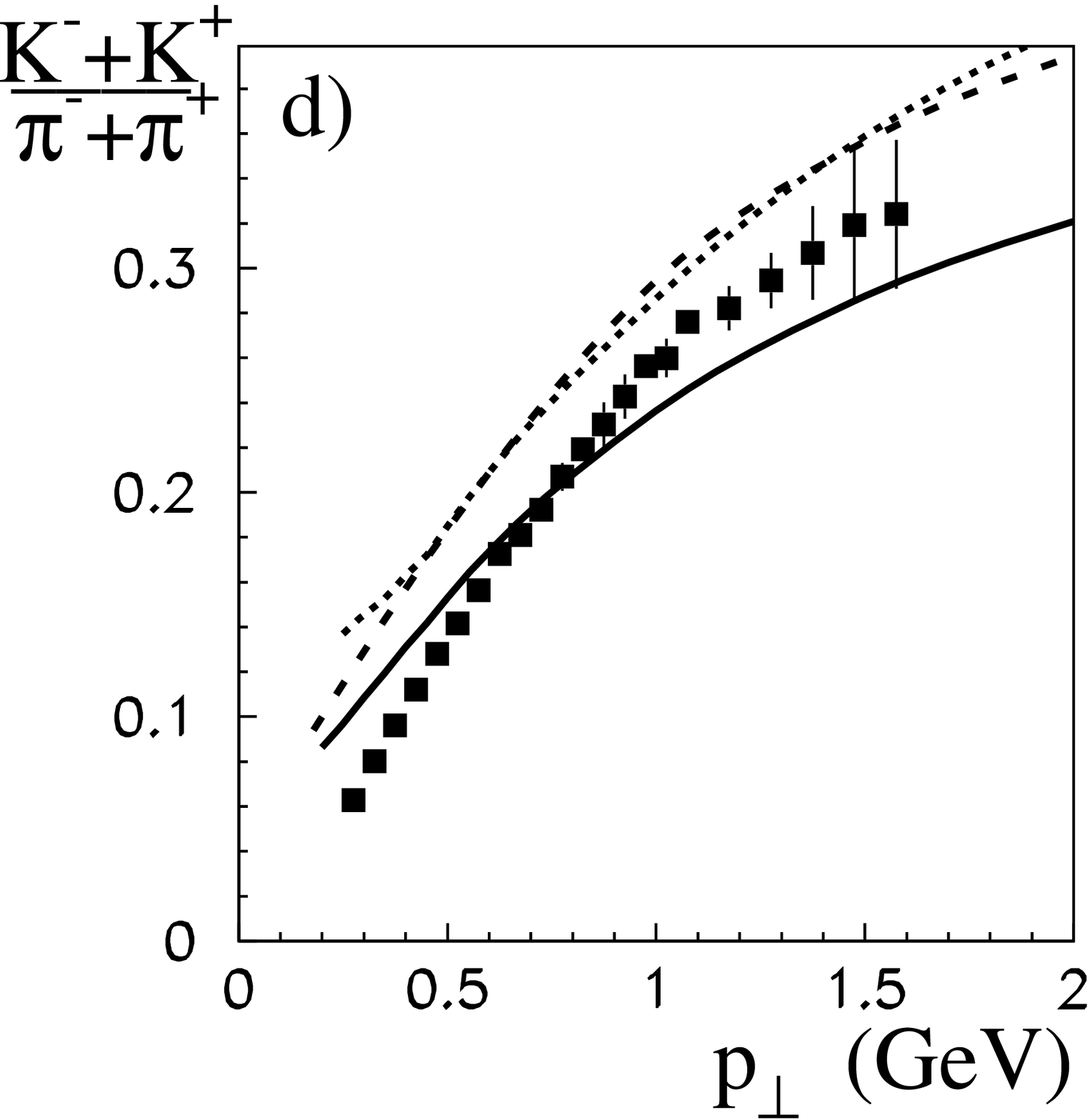} }
\caption{Identified particles ratios as a function of transverse momentum (transverse mass  for $\Omega / \Xi$)
for the our final strangeness and diquark suppressions (solid lines) in comparison with ALICE data from \cite{aliceA,aliceB,aliceE,aliceG}. The ratios without diquark (and strange diquark) suppression is shown by the dashed lines.
The results for Boltzmann statistics model are given also as a dotted lines.
\label{f5}}
\end{figure*}

The ALICE Collaboration published data showing also particle ratios as a function of particle transverse momentum. 
Taking into account that the modification of the statistics of the multiparticle 
production process was developed primarily for the transverse momentum description this kind of data could 
be valuable to verify the model. Comparison of our final model prediction and the data are shown in 
Fig.~\ref{f5}. The solid lines represent predictions of the discussed modified statistical model with the final suppressions as presented in the Tab.\ref{t4}, $\gamma_s = 0.66$ and both diquark suppression 
factors $\gamma_{qq}=\gamma_{ss}=0.5$.
The Boltzmann statistics results (dotted lines) 
are also given for a comparison. 
As it is seen in Fig.~\ref{f5}, the standard statistics does not work very well for the LHC ALICE data 
shown, as well as the the modified, Tsallis statistics without additional diquark suppression (dashed lines).
Only introducing the diquark suppression  effect with our chosen, first guess, values of suppression factors 
reproduces the data  better.
This is of course the effect of adding two new parameters and adjusting
the model to match the points, but the question if similar modification of the 
standard Boltzmann picture will give similar result is still open.

\section{Conclusions}
The modified thermodynamical model parameters found analyzing the transverse momentum distributions measured at 7 TeV
without any re-adjustment with the standard  strangeness suppression factor $\gamma$ of about 2/3 
and additional suppressions of diquark and strange diquark production were used for the calculations of identified particle multiplicities. 
We have shown that the introduction of non-extensive statistics to the
thermodynamical theory of multiparticle production in hadronic collisions opens an interesting possibility
%can give its consistent 
of the description of the hadronization process. 
%By the "consistency" we understand the
%The applicabilion of the same mechanism for the transverse momenta and the particle flavours creation description.

%The  can give the consistency of the model predictions with measurements on the average 
%particle ratios as well as their $p_\bot$ distributions.

\end{document}